\documentclass[preprintnumbers,superscriptaddress,showkeys,byrevtex]{revtex4}
\usepackage{amsmath,amsfonts,amssymb,amscd,amsxtra,amsthm}
\usepackage{graphicx}
\usepackage{epstopdf}
\usepackage{bm}
\usepackage{feynmp}
\usepackage{feynmp-auto}
\usepackage{orcidlink}
\usepackage{stackengine}
\usepackage{subcaption}
\usepackage{tikz-feynman}
\usepackage{pgfplots}
\usepgflibrary{plotmarks}
\usepackage{slashed}
\begin{document}
\preprint{PKNU-NuHaTh-2026}
%---------------------------------------------------------
\title{Zero-locus diagnostic of the direct contact term in $\phi\to\pi^+\pi^-\pi^0$}
%---------------------------------------------------------
\author{Seung-il Nam\,\orcidlink{0000-0001-9603-9775}}
\email{sinam@pknu.ac.kr}
\affiliation{Department of Physics, Pukyong National University (PKNU),  Busan 48513, Republic of Korea}
%---------------------------------------------------------
\author{Jung Keun Ahn\,\orcidlink{0000-0002-5795-2243}}
\affiliation{Department of Physics, Korea University, Seoul 02841, Republic of Korea}
%---------------------------------------------------------
\date{\today}
%---------------------------------------------------------
\begin{abstract}
We formulate, within a fixed FSI-improved amplitude convention, a zero-locus diagnostic for the direct contact term in $\phi\to\pi^+\pi^-\pi^0$.  The resonant $\rho\pi$ contribution is dressed channel by channel with the complex Omnès function, the direct term is multiplied by the averaged Omnès factor, and a real constant background is retained as a smooth residual amplitude.  In this convention, $\mathcal F^{\rm FSI}=R+c e^{i\delta_{\rm dir}}D$ with $c=g_{\phi3\pi}^{\rm eff}$, and the difference of two constrained invariant-mass gradients of the reduced Dalitz density gives a local quadratic identity for $c$ on the zero-contour.  The construction provides a closure and stability diagnostic complementary to a global Dalitz-plot fit.  For the benchmark input $g_{\phi3\pi}^{\rm eff}=-24.0~{\rm GeV}^{-3}$, we identify a numerically stable zero-contour branch around $M_{+0}=0.460\pm0.025~{\rm GeV}$ and $M_{+-}\le0.560~{\rm GeV}$, where the tight-contour median estimator reproduces $g_{\phi3\pi}^{\rm est}\simeq -24.0~{\rm GeV}^{-3}$, a closure recovery of the injected coupling within the chosen FSI-improved convention rather than an independent extraction.  We further vary the mass-window selection of the branch.  The median remains close to the injected value under shifts of the $M_{+0}$ window and moderate changes of its width, while the mean and standard deviation are more sensitive to outlier points near unstable quadratic branches.  The method therefore identifies Dalitz-plane regions where the direct-term interference is locally well-conditioned and provides useful guidance for control-region choices and systematic variations in future global fits.
\end{abstract}
%---------------------------------------------------------
\maketitle
%---------------------------------------------------------
\section{Introduction}
%---------------------------------------------------------
The decay $\phi\to\pi^+\pi^-\pi^0$ provides a useful setting in which the dominant $\rho\pi$ mechanism coexists with a much smaller direct contribution.  In chiral effective descriptions with vector mesons, the direct coupling $g_{\phi3\pi}$ is associated with the anomalous Wess-Zumino-Witten (WZW) sector~\cite{Wess:1971yu,Witten:1983tw,Rudaz:1984bz,Kaymakcalan:1983qq,Fujiwara:1984mp}, a contact-type contribution absent from the tree-level vector-meson-dominance (VMD) pole amplitude.  The event distribution is therefore shaped mainly by the three charge channels of the $\rho$ pole, while the direct term changes the density through interference rather than through a separately visible structure.  This feature makes the extraction of the contact contribution intrinsically convention dependent, because the apparent direct strength can be correlated with the adopted line shape, the normalization of the pole amplitude, the treatment of final-state interactions (FSI), and the residual background term.  Consequently, the direct component is usually determined through a global Dalitz-plot fit~\cite{KLOE:2003kas,ParticleDataGroup:2024cfk}.  Such fits are quantitatively powerful, but they do not directly identify the local Dalitz-plane regions most sensitive to the contact term.  Omnès-type dispersive representations~\cite{Niecknig:2012sj,Danilkin:2014cra} provide a systematic framework for precision amplitude studies by incorporating elastic rescattering in a controlled manner.  Here we address a narrower diagnostic question: whether local derivatives of the reduced Dalitz density can identify stable regions where the contact-term interference is resolved, and whether those regions can serve as a consistency check on the amplitude reconstruction.

The basic observation is that the difference between two constrained invariant-mass gradients of the reduced Dalitz density defines a nontrivial zero locus on the physical Dalitz surface.  This constraint is essential because only two invariant masses are independent, and the derivative must include the induced motion of the third invariant mass required by the Dalitz-surface sum rule.  On the exact zero locus, the direct coupling satisfies a local algebraic identity built from derivatives of the reference amplitude.  In the present FSI-improved convention, this identity is quadratic because the direct term carries an $s$-dependent complex Omnès factor that interferes with both the resonant amplitude and the residual background.  Because the zero locus is constructed from a density that already contains the input direct term, the identity is a closure relation rather than an independent measurement.  The useful output is instead a local map showing where the fitted density responds coherently to direct-contact interference and where the derivative reconstruction is sufficiently well-conditioned to serve as a control region.  This information is directly relevant for a global Dalitz analysis, since it indicates which bins or contour branches should be most sensitive to systematic changes in the direct term, the residual background, or the FSI convention.  We formulate this identity, identify the stability cuts required for a controlled implementation, and map out the construction's validity and failure modes within a self-contained FSI-improved amplitude.

A key distinction between the two approaches lies in their normalization conventions. The KLOE Dalitz-fit parametrization uses a dimensionless complex direct term, $A_{\rm dir}=a_d e^{i\phi_d}$, specified by a dimensionless magnitude and phase. In contrast, the present FSI-improved reduced amplitude employs a dimensionful effective direct coupling, so the coupling carries physical units and parametrizes the interaction strength within the present amplitude convention. As a result, the two parametrizations cannot be directly mapped onto each other without a complete refit within a unified amplitude convention. Therefore, any numerical comparison made below must be interpreted as indicative only, rather than as evidence for phenomenological tension with KLOE. Furthermore, the benchmark value used in the following analysis is not taken directly from KLOE experimental results, but is instead an injected input chosen specifically to test closure properties within the fixed convention of the present analysis. The paper is organized as follows. Section~II introduces the kinematics, the FSI-improved reference amplitude, and the reduced Dalitz density. Section~III derives the constrained gradient-difference zero-locus, the local quadratic identity, and the associated stability cuts. Section~IV presents the closure test and branch-stability analysis and discusses the data-level extension. Section~V summarizes the validity conditions and failure modes of the method as a diagnostic complementary to global and dispersive analyses.

%---------------------------------------------------------
\section{Framework}
%---------------------------------------------------------
We consider the decay process
%EQUATION>>>
\begin{equation}
\phi(p,\epsilon_\phi)\to\pi^+(p_+)\,\pi^-(p_-)\,\pi^0(p_0).
\end{equation}
%EQUATION>>>
The kinematics are described by the invariant masses
%EQUATION>>>
\begin{equation}
s_{+-}=(p_++p_-)^2,\quad s_{+0}=(p_++p_0)^2,\quad s_{-0}=(p_-+p_0)^2,
\label{eq:invariants}
\end{equation}
%EQUATION>>>
which satisfy the sum rule
%EQUATION>>>
\begin{equation}
s_{+-}+s_{+0}+s_{-0}=m_\phi^2+2m_{\pi^\pm}^2+m_{\pi^0}^2\equiv S_{\rm tot}.
\label{eq:sumrule}
\end{equation}
%EQUATION>>>
Using the PDG values $m_\phi = 1019.461~\mathrm{MeV}$, $m_{\pi^\pm}=139.570~\mathrm{MeV}$, and
$m_{\pi^0}=134.977~\mathrm{MeV}$~\cite{ParticleDataGroup:2024cfk}, the constant evaluates to
$S_{\rm tot}=1.09648~\mathrm{GeV}^2$. For comparison with the KLOE Dalitz analysis, we also use the variables
%EQUATION>>>
\begin{equation}
x=\frac{s_{-0}-s_{+0}}{2m_\phi},\quad y=\frac{(m_\phi-m_{\pi^0})^2-s_{+-}}{2m_\phi},
\label{eq:kloe_xy}
\end{equation}
%EQUATION>>>
which parametrize the physical phase-space region. These are dimensionful versions of the Dalitz coordinates. Experimental analyses typically use dimensionless coordinates obtained by dividing the energy differences by the available kinetic energy or an equivalent scale. The shapes shown below can therefore be compared with experimental Dalitz plots after a linear rescaling of axes, though the numerical values of $x$ and $y$ here should not be read off directly against the dimensionless KLOE variables. The kinematic boundary corresponds to $|x|\lesssim 0.30\,\text{GeV}$ and $0\leq y\lesssim 0.35\,\text{GeV}$.  The three $\rho$-pole bands occupy the loci at $s_{+-}=m_\rho^2$, $s_{+0}=m_\rho^2$, and $s_{-0}=m_\rho^2$, where $m_\rho=775.26\,\text{MeV}$~\cite{ParticleDataGroup:2024cfk}. In the variables of Eq.~(\ref{eq:kloe_xy}), these loci are
%EQUATION>>>
\begin{eqnarray}
s_{+-}&=&m_\rho^2,\quad y\simeq0.089\,\text{GeV},\cr
s_{+0}&=&m_\rho^2,\quad y-x\simeq0.4355\,\text{GeV},\cr
s_{-0}&=&m_\rho^2,\quad y+x\simeq0.4355\,\text{GeV}.
\label{eq:rho_bands}
\end{eqnarray}
%EQUATION>>>
The last two relations follow from the sum rule together with the definitions of $x$ and $y$. They are the diagonal $\rho^\pm$ bands in the two-dimensional Dalitz representation.  In the isospin-symmetric limit, charge reflection $p_+\leftrightarrow p_-$ gives $s_{+0}\leftrightarrow s_{-0}$, or equivalently $M_{+0}\leftrightarrow M_{-0}$ and $x\to -x$ at fixed $y$, so the two charge-reflected halves of the Dalitz plane carry equivalent branch information up to small charged-neutral mass effects and any explicit isospin-breaking terms.

The use of the invariant masses $M_{ij}=\sqrt{s_{ij}}$ rather than the squared variables is a practical choice for the derivative diagnostic.  It makes the resonance locations and branch windows appear on a directly interpretable mass scale and avoids the visual overemphasis on the upper part of the Dalitz plot that can occur when squared variables are used. The physical content is unchanged, provided that all derivatives are evaluated with the proper constrained Jacobian on the Dalitz surface.

In the $\phi$ rest frame, the decay amplitude is written as
%EQUATION>>>
\begin{equation}
\mathcal{M}_{\rm tot}(\lambda) =im_\phi\mathcal{F}(s_{+-},s_{+0},s_{-0})[\bm\epsilon^{(\lambda)}\cdot\bm n],
\label{eq:full_amp}
\end{equation}
%EQUATION>>>
where $\bm n\equiv\bm p_+\times\bm p_-$ and $\mathcal{F}$ denotes the reduced scalar amplitude.  At the tree level, we write
%EQUATION>>>
\begin{equation}
\mathcal{F}(s_{+-},s_{+0},s_{-0})
=\mathcal{F}_{\rho\pi}+\mathcal{F}_{\rm dir}+\mathcal{F}_{\rm BKG},
\label{eq:F_decomp}
\end{equation}
%EQUATION>>>
with
%EQUATION>>>
\begin{equation}
\mathcal{F}_{\rm dir}=g_{\phi3\pi},\qquad
\mathcal{F}_{\rho\pi}
=-2g_{\phi\rho\pi}g_{\rho\pi\pi}
\sum_{a=0,\pm}\frac{1}{D^{\rm GS}_{\rho_a}(s_a)},\qquad
\mathcal{F}_{\rm BKG}\in{\bf R}.
\label{eq:Ftree_terms}
\end{equation}
%EQUATION>>>
Here $s_0=s_{+-}$, $s_+=s_{+0}$, and $s_-=s_{-0}$.  This tree-level expression fixes the amplitude decomposition adopted throughout. The coupling $g_{\phi3\pi}$ in Eq.~(\ref{eq:Ftree_terms}) denotes the bare tree-level contact strength in this scheme, whereas $g_{\phi3\pi}^{\rm eff}$ introduced below is the effective strength after the FSI dressing and phase convention have been specified. In the numerical and derivative analyses below, the resonant and direct terms are dressed with the Omnès final-state interaction, and the background is kept as a real constant.

The separation in Eq.~(\ref{eq:Ftree_terms}) is a definition of the working scheme, not a unique physical decomposition.  In particular, a smooth real offset can partly compensate for changes in the pole normalization or in the treatment of higher-energy rescattering. The zero-locus test below is therefore applied only after this scheme has been fixed: it checks internal consistency, not a scheme-independent value of $g_{\phi3\pi}$.

The Gounaris-Sakurai (GS) propagator~\cite{Gounaris:1968mw} reads
%EQUATION>>>
\begin{equation}
D^{\rm GS}_\rho(s)=m_\rho^2-s+f_{\rm GS}(s)-i\sqrt{s}\Gamma_\rho(s),
\label{eq:GS_prop}
\end{equation}
%EQUATION>>>
with the energy-dependent width
%EQUATION>>>
\begin{equation}
\Gamma_\rho(s)=\Gamma_\rho\left[\frac{q(s)}{q(m_\rho^2)}\right]^{3}\frac{m_\rho}{\sqrt{s}},
\quad  q(s)=\tfrac{1}{2}\sqrt{s-4m_{\pi^\pm}^2} ,
\label{eq:GS_width}
\end{equation}
and the real dispersive correction
\begin{equation}
f_{\rm GS}(s)=\frac{\Gamma_\rho m_\rho^2}{\pi q_0^3}\left[q^2(s)\bigl(h(s)-h_0\bigr)
+q_0^2(m_\rho^2-s)\,\dot{h}_0\right],
\label{eq:GS_fGS}
\end{equation}
%EQUATION>>>
where
%EQUATION>>>
\begin{equation}
h(s)=\frac{2q(s)}{\pi\sqrt{s}}\ln\frac{\sqrt{s}+2q(s)}{2m_{\pi^\pm}},\quad 
\dot{h}_0=\frac{h(m_\rho^2)}{8q_0^2}+\frac{1}{\pi m_\rho^2}.
\label{eq:GS_h}
\end{equation}
%EQUATION>>>
With $\Gamma_\rho=149.1\,\text{MeV}$~\cite{ParticleDataGroup:2024cfk}, the key numerical values are $q_0=0.3616\,\text{GeV}$ and $h_0=0.499$. For simplicity, we use the charged-pion mass in the width function for all three $\rho$ channels. As a quantitative check of this approximation, replacing the equal-mass momentum in the charged $\rho^\pm\to\pi^\pm\pi^0$ width by the exact two-body momentum changes $\Gamma_\rho(s)$ by about $1.7$--$2.8\%$ across the selected low-mass interval $M_{+0}=0.435$--$0.485\,\text{GeV}$. When only this width-function mass assignment is varied, the corresponding change in $|1/D_\rho^{\rm GS}|$ is below $0.01\%$ in that interval and remains below about $0.1\%$ for the charge-reflected high-mass channel. This approximation is therefore numerically negligible for the present closure diagnostic, though the exact charged-neutral masses should be restored in a precision fit. In the neutral channel, a localized contribution from $\rho^0$-$\omega$ mixing may further be included~\cite{McNamee:1974vb,Coon:1987kt,Bernicha:1994re,OConnell:1995nse,Gardner:1997ie}. What matters for the construction below is the explicit separation between the Dalitz-dependent resonant term and the constant direct contribution.

After averaging over the initial $\phi$ polarizations, the squared amplitude becomes
%EQUATION>>>
\begin{equation}
\overline{|\mathcal{M}_{\rm tot}|^2}=\frac{1}{3}m_\phi^2|\bm n|^2|\mathcal{F}(s_{+-},s_{+0},s_{-0})|^2.
\label{eq:Msqred}
\end{equation}
%EQUATION>>>
The elastic $I=1$, $P$-wave final-state interaction is incorporated through the Omnès function
\begin{equation}
\Omega_1(s)=
\exp\left[
\frac{s}{\pi}\int_{4m_\pi^2}^{\Lambda_\Omega^2}ds'\,
\frac{\delta_1^1(s')}{s'(s'-s-i\epsilon)}
\right],
\qquad
\Omega_1(s+i0)=|\Omega_1(s)|e^{i\delta_1^1(s)} .
\label{eq:omnes_def}
\end{equation}
Here $s+i0$ denotes the boundary value approached from the upper rim of the physical cut, i.e. $s+i\epsilon$ with $\epsilon\to0^+$, and $\delta_1^1(s)$ is the elastic $\pi\pi$ phase shift in the $I=1$, $L=1$ ($P$-wave) channel.
For the three two-pion channels, we define
\begin{equation}
\Omega_0=\Omega_1(s_{+-}),\qquad
\Omega_+=\Omega_1(s_{+0}),\qquad
\Omega_-=\Omega_1(s_{-0}),
\label{eq:omega_channels}
\end{equation}
and dress the direct term with the averaged factor
\begin{equation}
\Omega_{\rm dir}(s_{+-},s_{+0},s_{-0})
=\frac{\Omega_0+\Omega_++\Omega_-}{3}.
\label{eq:Omega_dir}
\end{equation}
To avoid double-counting the elastic rescattering already contained in the empirical $\rho\to\pi\pi$ width, the Omnès-dressed exchange amplitude uses
\begin{equation}
g_{\rho\pi\pi}^{\rm bare}
=\frac{g_{\rho\pi\pi}^{\rm phys}}{|\Omega_1(m_\rho^2)|}.
\label{eq:grhopipi_bare}
\end{equation}
This prescription keeps the normalization of the exchange amplitude tied to the physical $\rho\to\pi\pi$ coupling while allowing the Omnès factor to supply the channel-dependent phase and modulus variation away from the pole.  The direct term is treated differently. It is multiplied by the averaged factor in Eq.~(\ref{eq:Omega_dir}) because it represents a three-pion contact production structure rather than a single two-pion pole channel.  This difference is the origin of the quadratic term in the estimator derived below.
The FSI-improved reduced amplitude is then
\begin{equation}
\mathcal F^{\rm FSI}
=g_{\phi3\pi}^{\rm eff}e^{i\delta_{\rm dir}}\Omega_{\rm dir}
-2g_{\phi\rho\pi}^{\rm eff}g_{\rho\pi\pi}^{\rm bare}
\sum_{a=0,\pm}
\frac{\Omega_a(s_a)}{D^{\rm GS}_{\rho_a}(s_a)}
+\mathcal F_{\rm BKG}.
\label{eq:FFSI_full}
\end{equation}
For the neutral channel, $\rho^0$-$\omega$ mixing is included through
\begin{equation}
\frac{\Omega_0}{D^{\rm GS}_{\rho^0}(s_{+-})}
\to
\frac{\Omega_0}{D^{\rm GS}_{\rho^0}(s_{+-})}
\left[
1+\delta_{\rho\omega}\frac{m_\omega^2}{D_\omega(s_{+-})}
\right].
\label{eq:mixing_fsi}
\end{equation}
The background term $\mathcal F_{\rm BKG}$ is left undressed and retained only as a residual real constant amplitude. This separation matters for the diagnostic below: the direct term carries the averaged elastic FSI factor associated with the three-body contact production amplitude, whereas the background serves as a smooth real offset absorbing residual strength not captured by the minimal resonant-plus-direct ansatz. Its value should therefore not be interpreted as a separate physical contact coupling.  We have verified that the selected branch is not defined by the absolute value of $\mathcal F_{\rm BKG}$ alone, although its location and the conditioning of the quadratic roots change when the smooth offset is varied. A full profiling over $\mathcal F_{\rm BKG}$ is part of the global-fit implementation. The numerical reference setup used in the present closure test is
\begin{equation}
g_{\phi\rho\pi}^{\rm eff}=6.27~{\rm GeV}^{-1},\quad
g_{\phi3\pi}^{\rm eff}=-24.0~{\rm GeV}^{-3},\quad
\delta_{\rm dir}=-25^\circ,\quad
\mathcal F_{\rm BKG}=-50.0~{\rm GeV}^{-3},\quad
\Lambda_\Omega=1.60~{\rm GeV}.
\label{eq:reference_inputs}
\end{equation}
These are phenomenological effective parameters fixing the FSI-improved amplitude used throughout, not values derived from the zero-locus construction itself.  As discussed in Sec.~I, $g_{\phi3\pi}^{\rm eff}=-24.0~{\rm GeV}^{-3}$ is a benchmark input injected to test closure, motivated by the Dalitz-density setup of Ref.~\cite{Nam:2026dmr} but not identified with it or with the KLOE parametrization. The diagnostic construction below does not rely on Ref.~\cite{Nam:2026dmr} as an external input. The present work uses this value only as a closure benchmark and does not import a fitted direct-coupling result from Ref.~\cite{Nam:2026dmr}.

This motivates the definition of the FSI-improved reduced Dalitz density,
%EQUATION>>>
\begin{equation}
\rho_{\rm red}^{\rm FSI}(s_{+-},s_{+0},s_{-0})
\equiv|\mathcal{F}^{\rm FSI}(s_{+-},s_{+0},s_{-0})|^2.
\label{eq:rho_red_def}
\end{equation}
%EQUATION>>>
For the gradient construction, it is useful to isolate the unknown direct strength by writing
\begin{equation}
\mathcal F^{\rm FSI}=R+c\,e^{i\delta_{\rm dir}}D,\qquad
c\equiv g_{\phi3\pi}^{\rm eff},\quad
D\equiv\Omega_{\rm dir},\quad
R\equiv\mathcal F_{\rho\pi}^{\rm FSI}+\mathcal F_{\rm BKG}.
\label{eq:R_D_def}
\end{equation}
Here, $R$ contains the resonant exchange amplitude and the smooth background after the FSI convention has been fixed, whereas $c$ multiplies only the direct production structure.  The phase $\delta_{\rm dir}$ is kept explicit so that the fitted sign and magnitude of $c$ are not confused with a redefinition of the complex Omnès factor.  With this notation, changes in the density caused by the direct term appear through two pieces: a term linear in $c$, which is the interference with $R$, and a term quadratic in $c$, which is the direct contribution squared.
Then
\begin{equation}
\rho_{\rm red}^{\rm FSI}
=|R|^2
+2c\,{\rm Re}\!\left[e^{i\delta_{\rm dir}}D R^*\right]
+c^2|D|^2 .
\label{eq:rho_red_expand}
\end{equation}
The zero-locus identity is thus quadratic in the direct effective coupling. A linearized identity would be recovered only in the special limit $D=1$, $\delta_{\rm dir}=0$, $\mathcal F_{\rm BKG}=0$, and $\Delta |D|^2=0$.

%---------------------------------------------------------
\section{Gradient-difference construction}
%---------------------------------------------------------
We define the gradient-difference quantity
%EQUATION>>>
\begin{equation}
\Delta_G(M_{+-},M_{+0}) \equiv
\frac{d \rho_{\mathrm{red}}^{\rm FSI}}{d M_{+-}}-\frac{d \rho_{\mathrm{red}}^{\rm FSI}}{d M_{+0}},\quad M_{ij}\equiv \sqrt{s_{ij}}.
\label{eq:DeltaG}
\end{equation}
%EQUATION>>>
The derivatives are constrained derivatives on the physical Dalitz surface. With $(M_{+-},M_{+0})$ as independent coordinates, the third invariant mass is fixed by
%EQUATION>>>
\begin{equation}
M_{-0}=\sqrt{S_{\rm tot}-M_{+-}^2-M_{+0}^2},
\label{eq:Mminus0_constraint}
\end{equation}
%EQUATION>>>
and the induced variation of $M_{-0}$ must be included as
%EQUATION>>>
\begin{eqnarray}
\frac{d}{dM_i}=\frac{\partial}{\partial M_i}-\frac{M_i}{M_{-0}}\frac{\partial}{\partial M_{-0}},\quad i=+-,+0 .
\end{eqnarray}
%EQUATION>>>
Since the three invariant masses are not mutually independent, this constraint also fixes the meaning of the zero-locus condition below.  A derivative taken at fixed $M_{-0}$ would correspond to a displacement outside the physical Dalitz surface and would generate a different curve.  Eq.~(\ref{eq:Mminus0_constraint}) is therefore part of the definition of the diagnostic, not merely a numerical implementation detail.

The choice in Eq.~(\ref{eq:DeltaG}) compares the neutral-channel direction with one charged-channel direction and makes a near-vertical branch of the zero locus transparent in the $(M_{+-}, M_{+0})$ plane. The cyclic alternatives $d/dM_{+0}-d/dM_{-0}$ and $d/dM_{-0}-d/dM_{+-}$ give equivalent constructions up to a permutation of the Dalitz variables and a corresponding relabeling of the selected branch.  As a numerical cross-check, we repeated the branch search with the two cyclic choices using the same pole vetoes and contour tolerance.  The selected branches are mapped into the corresponding charge-reflected regions, and the tight-contour median of $g_{\phi3\pi}^{\rm est}$ changes by less than the quoted branch-window spread.  We therefore use Eq.~(\ref{eq:DeltaG}) as the default orientation because it gives the clearest visual separation of the low-mass branch used below. The condition
%EQUATION>>>

\begin{equation}
\Delta_G(M_{+-},M_{+0})=0
\label{eq:locus_condition}
\end{equation}
%EQUATION>>>
defines a locus on the physical Dalitz plane. At regular points where $\nabla\Delta_G\neq0$, the zero set is locally a one-dimensional contour. Multiple connected components, endpoints, or branch mergers can occur near critical points with $\nabla\Delta_G=0$. We do not assume global uniqueness of the zero locus, and instead select one regular, numerically stable branch as a controlled closure-test region. This branch is not selected by optimizing the recovered value of $g_{\phi3\pi}^{\rm eff}$, but by the geometric regularity of the zero contour and the stability criteria in Eq.~(\ref{eq:eps_dmin_cuts}).

Substituting Eq.~(\ref{eq:rho_red_expand}) into Eq.~(\ref{eq:DeltaG}), and denoting
\begin{equation}
\Delta\equiv\frac{d}{dM_{+-}}-\frac{d}{dM_{+0}},
\label{eq:Delta_op}
\end{equation}
one obtains on the zero locus
%EQUATION>>>
\begin{equation}
A+2cB+c^2C=0,
\label{eq:locus_expand}
\end{equation}
%EQUATION>>>
where
\begin{equation}
A=\Delta |R|^2,\qquad
B=\Delta\,{\rm Re}\!\left[e^{i\delta_{\rm dir}}D R^*\right],\qquad
C=\Delta |D|^2 .
\label{eq:ABC_def}
\end{equation}
On the exact zero locus defined by Eq.~(\ref{eq:locus_condition}), the direct effective coupling therefore satisfies the local quadratic identity
%EQUATION>>>
\begin{equation}
g_{\phi3\pi}^{\rm est}(M_{+-},M_{+0})
=\frac{-B\pm\sqrt{B^2-AC}}{C}.
\label{eq:cest}
\end{equation}
%EQUATION>>>
The physical branch is chosen continuously from the reference value in Eq.~(\ref{eq:reference_inputs}).  If $|C|$ is numerically small, the stable limiting form is
\begin{equation}
g_{\phi3\pi}^{\rm est}\simeq -\frac{A}{2B}.
\label{eq:cest_linear_limit}
\end{equation}
Eq.~(\ref{eq:cest}) is the central algebraic result, expressing the direct effective coupling through derivatives of the FSI-improved reference amplitude, including the real background and the averaged Omnès dressing of the direct term. Away from the zero locus, the same expression can be plotted as a diagnostic variable but carries no meaning as a local coupling measurement.

The quadratic form also clarifies three sources of instability. First, the discriminant $\mathcal D_c=B^2-AC$ must stay non-negative for a real local estimator. When both $B$ and $C$ become small, the two roots grow extremely sensitive to small numerical changes in the derivative fields, and the two algebraic branches may exchange their relative proximity to the reference value near turning points of the zero contour. The stability cuts introduced below are accordingly numerical regularity conditions, not physical cuts on the decay amplitude, that select the portion of the zero locus where the diagnostic is meaningful. In an experimental implementation, the zero locus would have to be reconstructed from an efficiency-corrected Dalitz density, with the reference amplitude entering $A$, $B$, and $C$, which are fixed independently by a global fit or a dispersive representation.

For numerical work, we apply the estimator with explicit stability cuts.  We define the quadratic discriminant and denominator controls as
%EQUATION>>>
\begin{equation}
\mathcal D_c\equiv B^2-AC,\qquad
B_{\rm dir}(M_{+-},M_{+0})\equiv B,\qquad
C_{\rm dir}(M_{+-},M_{+0})\equiv C,
\label{eq:denom_def}
\end{equation}
%EQUATION>>>
and retain only points satisfying
%EQUATION>>>
\begin{equation}
|\Delta_G|<\varepsilon_G,\qquad
\mathcal D_c\ge0,\qquad
|B_{\rm dir}|>B_{\rm min}
\quad\hbox{or}\quad
|C_{\rm dir}|>C_{\rm min}.
\label{eq:eps_dmin_cuts}
\end{equation}
%EQUATION>>>
With the present normalization, the reduced amplitude has mass dimension $[\mathcal F]=\mathrm{GeV}^{-3}$, so $[|\mathcal F|^2]=\mathrm{GeV}^{-6}$ and $[\Delta_G]=\mathrm{GeV}^{-7}$.  In Eq.~(\ref{eq:ABC_def}), $A$ has dimension $\mathrm{GeV}^{-7}$, $B$ has dimension $\mathrm{GeV}^{-4}$, and $C$ has dimension $\mathrm{GeV}^{-1}$, fixing the units of $\varepsilon_G$, $B_{\rm min}$, and $C_{\rm min}$ in Table~\ref{tab:inputs}. The first condition in Eq.~(\ref{eq:eps_dmin_cuts}) restricts the sample to the zero locus within a finite numerical tolerance, while the remaining conditions remove branch points where the quadratic estimator becomes unstable. These thresholds are varied below to test stability rather than being treated as fixed physical inputs.

Three further cuts complete the selection. Points too close to a $\rho$ band are removed using $\delta_\rho=45~\mathrm{MeV}$, i.e. $|M_{ij}-m_\rho|>\delta_\rho$ for the channels in the branch sample; this excludes regions where the derivative field is dominated by the steep pole structure rather than by the smooth interference pattern. In the neutral channel we additionally require $|M_{+-}-m_\omega|>\delta_\omega$ with $m_\omega=782.66~\mathrm{MeV}$ and $\delta_\omega\simeq8.7~\mathrm{MeV}$~\cite{ParticleDataGroup:2024cfk}, so that the localized $\rho^0$-$\omega$ deformation is not mistaken for a smooth direct-term signal. Among the surviving branches of the zero locus we then pick out the low-mass one around $M_{+0}=0.460\pm0.025~\mathrm{GeV}$ with $M_{+-}\le0.560~\mathrm{GeV}$, where Eq.~(\ref{eq:eps_dmin_cuts}) acts as a contour-restricted closure test. The numerical inputs and branch-selection settings are summarized in Table~\ref{tab:inputs}.
%---------------------------------------------------------
\begin{table}[b]\centering
\setlength{\tabcolsep}{10pt}
\begin{tabular}{ccc|ccc}\hline\hline
\multicolumn{3}{c|}{Kinematic and amplitude inputs} &\multicolumn{3}{c}{Branch and stability settings}\\
\hline
$m_{\phi}$ & $1019.461~\mathrm{MeV}$ & \hspace{0.5em}PDG &
$m_{\omega}$ & $782.66~\mathrm{MeV}$ & PDG\\
$m_{\pi^\pm}$ & $139.570~\mathrm{MeV}$ & PDG &
$\delta_{\rho}$ & $45~\mathrm{MeV}$ & \hspace{0.5em}branch cut\\
$m_{\pi^0}$ & $134.977~\mathrm{MeV}$ & PDG &
$\delta_{\omega}$ & $\simeq8.7~\mathrm{MeV}$ & \hspace{0.5em}veto\\
$S_{\rm tot}$ & $1.09648~\mathrm{GeV}^2$ & \hspace{0.5em}Eq.~(\ref{eq:sumrule}) &
$g_{\rho\pi\pi}$ & $5.98$ & \hspace{0.5em}input\\
$m_{\rho}$ & $775.26~\mathrm{MeV}$ & PDG &
$g_{\phi\rho\pi}^{\rm eff}$ & $6.27~\mathrm{GeV}^{-1}$ & \hspace{0.5em}reference\\
$\Gamma_{\rho}$ & $149.1~\mathrm{MeV}$ & PDG &
$g_{\phi3\pi}^{\rm eff}$ & $-24.0~\mathrm{GeV}^{-3}$ & \hspace{0.5em}reference\\
$q_0$ & $0.3616~\mathrm{GeV}$      & \hspace{0.5em}Eq.~(\ref{eq:GS_width}) &
$M_{+0}^{\rm br}$ & $0.460\pm0.025~\mathrm{GeV}$ & \hspace{0.5em}branch\\
$h_0$ & $0.499$ & \hspace{0.5em}Eq.~(\ref{eq:GS_h}) & $M_{+-}^{\rm cut}$ & $\le0.560~\mathrm{GeV}$ & \hspace{0.5em}low side\\
$\varepsilon_G$ & $50$--$2000~\mathrm{GeV}^{-7}$ & \hspace{0.5em}scan &
$B_{\rm min},C_{\rm min}$ & \hspace{0.5em}scan & \hspace{0.5em}stability\\
\hline\hline
\end{tabular}
\caption{Input parameters and branch-selection settings used in the gradient-difference analysis. Masses and widths are taken from PDG~\cite{ParticleDataGroup:2024cfk}. The remaining entries define the FSI-improved reference amplitude and the branch/cut settings.}
\label{tab:inputs}
\end{table}
%---------------------------------------------------------

%---------------------------------------------------------
\section{Results and discussion}
\begin{figure}[t]
\topinset{(a)}{\includegraphics[width=7.5cm]{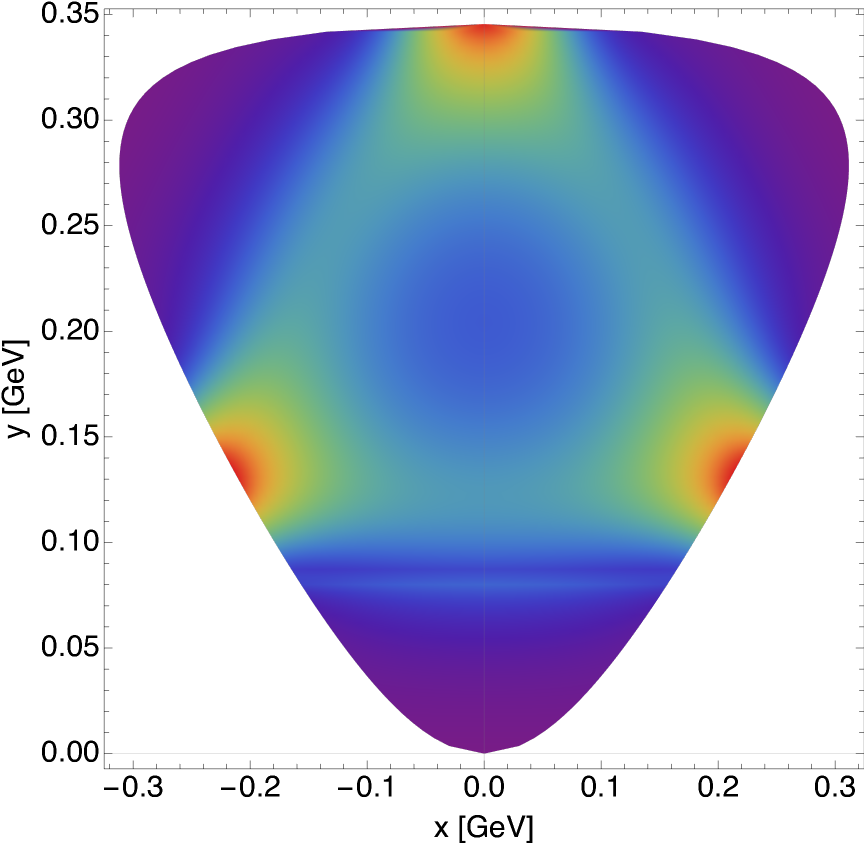}}{-0.4cm}{0.0cm}
\topinset{(b)}{\includegraphics[width=7.5cm]{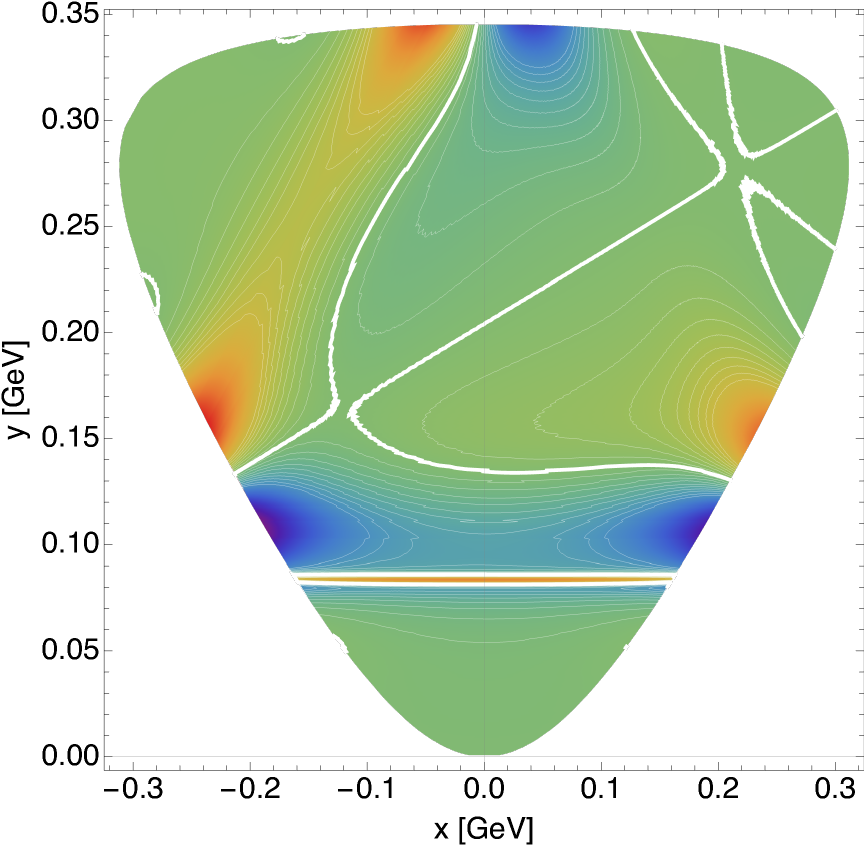}}{-0.4cm}{0.0cm}
\caption{(Color online) (a)~$\rho_{\rm red}^{\rm FSI}$ in Eq.~(\ref{eq:rho_red_def}) as a function of  $x$ and $y$ for the FSI-improved reference amplitude.  (b)~Gradient-difference map $\Delta_G$ in Eq.~(\ref{eq:DeltaG}). The thick curve is the zero-contour $\Delta_G=0$.}
\label{FIG1}
\end{figure}

%---------------------------------------------------------
The reduced Dalitz density $\rho_{\rm red}^{\rm FSI}$ and the gradient-difference map $\Delta_G$ are shown in Fig.~\ref{FIG1} as functions of $x$ and $y$, computed with the FSI-improved parameter set in Eq.~(\ref{eq:reference_inputs}).  Fig.~\ref{FIG1}(a) reproduces the expected three-band peaking structure associated with the $\rho^0\pi^0$, $\rho^+\pi^-$, and $\rho^-\pi^+$ mechanisms.  The direct and background terms do not generate separate bands. Instead, they change the density's height and local slope by interfering with the resonant amplitude.  This motivates the use of a derivative diagnostic, since the direct contribution is not visually isolated in the Dalitz density but can still leave a coherent imprint in the slope field of the reduced distribution.

Fig.~\ref{FIG1}(b) shows the gradient-difference field together with the extracted zero-contour $\Delta_G=0$, the geometric object on which Eq.~(\ref{eq:cest}) is defined.  Near resonance bands the slope field changes too rapidly, and near quadratic branch points the local root in Eq.~(\ref{eq:cest}) becomes ill-behaved, so only selected parts of the contour are suitable for a closure test.  The branch used below is picked for being regular and numerically robust under moderate changes of the branch window, not for any claim of uniqueness.
%---------------------------------------------------------
%---------------------------------------------------------

\begin{figure}[t]
\topinset{(a)}{\includegraphics[width=7.5cm]{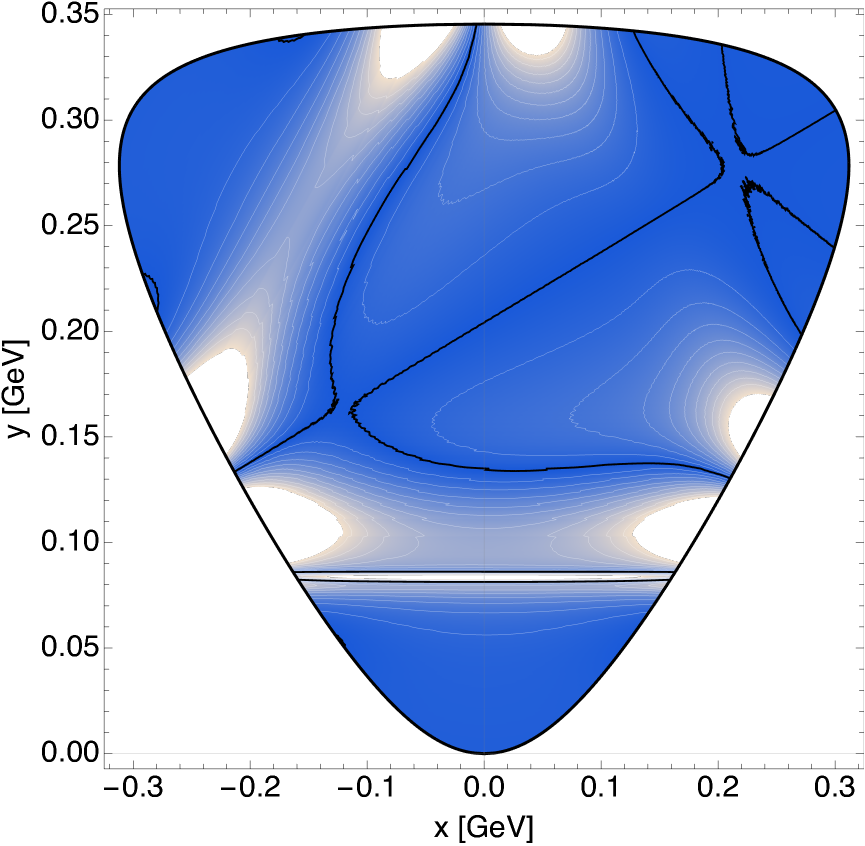}}{-0.4cm}{0.0cm}
\topinset{(b)}{\includegraphics[width=7.5cm]{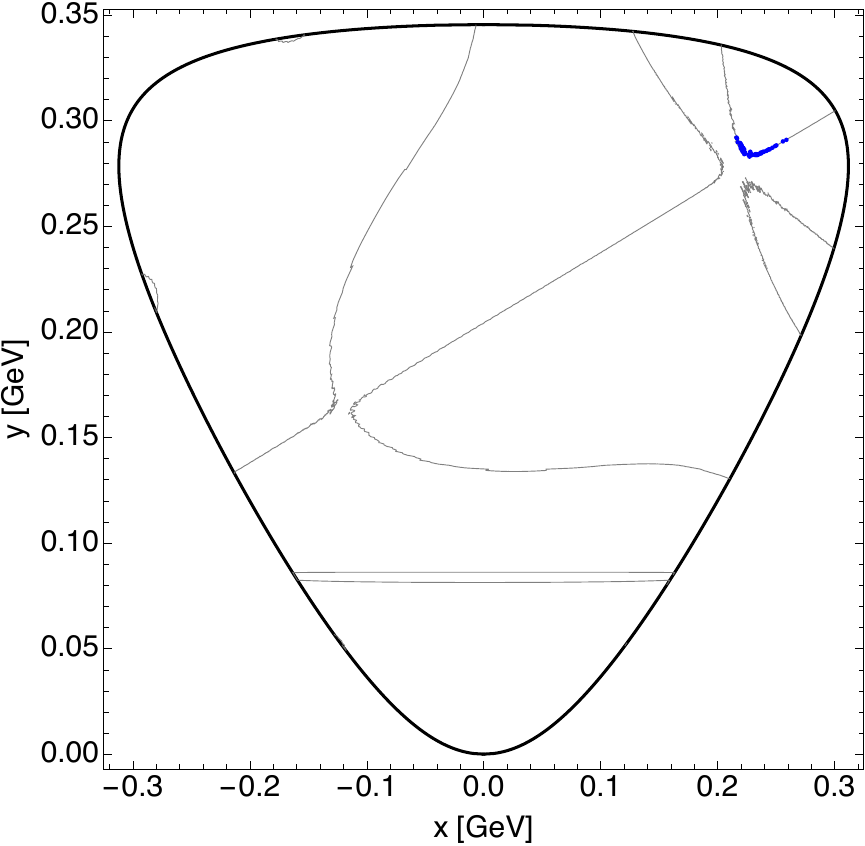}}{-0.4cm}{0.0cm}
 \caption{(Color online) (a) Pointwise diagnostic ratio $g_{\phi3\pi}^{\rm est}(M_{+-},M_{+0})$ from Eq.~(\ref{eq:cest}) in the FSI-improved convention. Large fluctuations originate from the $\rho$-pole bands and from small values of the quadratic stability factors. The displayed color range is saturated to keep the regular branch visible. The color map is shown only to visualize the algebraic instability structure; estimator values away from the selected zero-contour branch should not be interpreted as local coupling values. It is not an extraction map, but a visualization of the algebraic conditioning of Eq.~(\ref{eq:cest}). (b) Selected branch-strip region around $M_{+0}=0.460\pm0.025~\mathrm{GeV}$ and $M_{+-}\le0.560~\mathrm{GeV}$. This strip is used only as a contour-restricted closure-test region.}
\label{FIG2}
\end{figure}

The pointwise diagnostic ratio $g_{\phi3\pi}^{\rm est}(M_{+-}, M_{+0})$ from Eq.~(\ref{eq:cest}) is displayed in Fig.~\ref{FIG2}.  Fig.~\ref{FIG2}(a) shows the diagnostic map over the displayed Dalitz region. Large excursions appear near the $\rho$ bands and near points where the quadratic stability factors become small, confirming that the ratio is not a stable observable over the full Dalitz plane.  In producing the map, the color scale is saturated outside the displayed range so that the regular branch can be seen. Unsaturated outliers occur only in the unstable regions rejected by Eq.~(\ref{eq:eps_dmin_cuts}).

Fig.~\ref{FIG2}(b) shows the stable branch strip selected from the zero-contour calculation,
\begin{equation}
M_{+0}=0.460\pm0.025~\mathrm{GeV},\qquad M_{+-}\le0.560~\mathrm{GeV}.
\end{equation}
This strip is a closure-test region only: its location follows from the zero-contour geometry of the FSI-improved amplitude rather than from resonance kinematics, and carries no implication of a new physical structure. It is simply a part of the Dalitz surface where the derivative balance in Eq.~(\ref{eq:locus_expand}) is well conditioned enough to test whether the inserted direct effective coupling can be recovered.  The primary closure scan is discussed first, followed by a separate branch-window check as a supplementary test.

%---------------------------------------------------------

%---------------------------------------------------------
%---------------------------------------------------------

Applying Eq.~(\ref{eq:eps_dmin_cuts}) directly to the zero locus gives the disciplined version of the method. Table~\ref{tab:stability} gives a representative scan on the branch used in Fig.~\ref{FIG2}(b), obtained from a dense numerical grid.  Here and in the following tables, $N$ denotes the number of accepted grid points after the zero-contour, pole-veto, branch-window, and quadratic-regularity cuts have all been applied.  We checked the calculation by changing the mass-grid spacing in the contour reconstruction by a factor of about two and by repeating the derivative evaluation with the corresponding finite-difference step: the tight-contour median changes only at the sub-$0.1~\mathrm{GeV}^{-3}$ level, while the width and the mean shift more visibly as additional near-branch-point grid points enter the sample. The median's robustness therefore reflects a property of the regular zero-locus branch itself, and the growth of the mean and width with looser cuts is a quadratic-branch instability rather than a discretization artifact. For $\varepsilon_G=100~\mathrm{GeV}^{-7}$ the selected branch gives
\begin{equation}
\mathrm{median}\left[g_{\phi3\pi}^{\rm est}\right]
=-23.984~\mathrm{GeV}^{-3},
\end{equation}
to be compared with the input value $g_{\phi3\pi}^{\rm eff}=-24.0~\mathrm{GeV}^{-3}$.  Table~\ref{tab:stability} traces what happens as $\varepsilon_G$ is relaxed: the mean and width grow steadily, driven by a small but increasing number of points near unstable quadratic branches, while the median drifts only slowly. The $\varepsilon_G=2000~\mathrm{GeV}^{-7}$ row marks the point where this growth overtakes the median itself, and we use it here only to illustrate where the diagnostic breaks down; the rows at $\varepsilon_G\lesssim200~\mathrm{GeV}^{-7}$ define the region we treat as a genuine closure test.

%---------------------------------------------------------
\begin{table}[b]\centering
\setlength{\tabcolsep}{10pt}
\begin{tabular}{ccccc}\hline\hline
$\varepsilon_G~[\mathrm{GeV}^{-7}]$ & $N$ &
$\langle g_{\phi3\pi}^{\rm est}\rangle~[\mathrm{GeV}^{-3}]$ &
median $[\mathrm{GeV}^{-3}]$ & $\sigma~[\mathrm{GeV}^{-3}]$\\
\hline
50   & 31  & $-23.86$ & $-24.02$ & $0.67$\\
100  & 52  & $-23.11$ & $-23.98$ & $5.13$\\
200  & 111 & $-22.97$ & $-23.99$ & $7.41$\\
500  & 239 & $-21.56$ & $-23.94$ & $14.78$\\
1000 & 426 & $-26.72$ & $-24.15$ & $31.37$\\
2000 & 868 & $-35.56$ & $-27.47$ & $30.23$\\
\hline\hline
\end{tabular}
\caption{Branch-centered closure stability scan for the FSI-improved amplitude. The branch window is $M_{+0}=0.460\pm0.025~\mathrm{GeV}$ and $M_{+-}\le0.560~\mathrm{GeV}$. The median is the robust closure indicator in the tight-contour region. The mean and width reveal the outlier sensitivity of the pointwise quadratic estimator as the contour tolerance is relaxed. The largest-tolerance row illustrates the breakdown of the diagnostic window and is not used as an extraction window.}
\label{tab:stability}
\end{table}
%---------------------------------------------------------

As a second check, we vary the mass-window definition of the selected branch at fixed $\varepsilon_G=100~\mathrm{GeV}^{-7}$, to test whether the closure is tied to a narrowly chosen strip rather than being optimized for it a posteriori. We first vary the half-width of the $M_{+0}$ strip, then shift its center, and finally change the upper cut on $M_{+-}$, keeping the other two settings fixed in each case. The accepted points in these scans have already passed the zero-contour, pole-veto, $\omega$-veto, and quadratic-regularity cuts.

Table~\ref{tab:branch_sensitivity} varies the half-width $\delta M_{+0}$ of the branch window centered at $M_{+0}=0.460~\mathrm{GeV}$.  The median remains close to $-24~\mathrm{GeV}^{-3}$ throughout the scan.  The mean and standard deviation, however, grow when the window is widened beyond about $20~\mathrm{MeV}$, indicating that the wider strip begins to include points near less well-conditioned parts of the quadratic root.  This behavior is useful diagnostically: the stable information is carried by the median of the contour-restricted branch, whereas the mean is a sensitive monitor of residual outliers.

%---------------------------------------------------------
\begin{table}[t]\centering
\setlength{\tabcolsep}{9pt}
\begin{tabular}{ccccc}\hline\hline
$\delta M_{+0}~[\mathrm{MeV}]$ &
$N$ &
$\langle g_{\phi3\pi}^{\rm est}\rangle$ &
median &
$\sigma$\\
\hline
15 & 40 & $-23.741$ & $-24.016$ & $1.05$\\
20 & 47 & $-23.581$ & $-24.011$ & $1.42$\\
25 & 52 & $-23.108$ & $-23.984$ & $5.13$\\
30 & 56 & $-22.348$ & $-23.958$ & $6.54$\\
35 & 65 & $-22.503$ & $-23.965$ & $7.21$\\
\hline\hline
\end{tabular}
\caption{Branch half-width sensitivity at fixed $\varepsilon_G=100~\mathrm{GeV}^{-7}$, $\delta_\rho=45~\mathrm{MeV}$, $M_{+0}^{\rm br}=0.460~\mathrm{GeV}$, and $M_{+-}\le0.560~\mathrm{GeV}$. All estimator values are in $\mathrm{GeV}^{-3}$. The median is stable near the injected value, while the mean and width reveal increasing outlier sensitivity as the branch window is enlarged.}
\label{tab:branch_sensitivity}
\end{table}
%---------------------------------------------------------

Table~\ref{tab:branch_center_sensitivity} shifts the center of the branch window while keeping $\delta M_{+0}=25~\mathrm{MeV}$ and $M_{+-}\le0.560~\mathrm{GeV}$.  Again, the median stays close to the injected value over the whole range.  The lower-center choices, especially $M_{+0}^{\rm br}=0.450$--$0.455~\mathrm{GeV}$, show larger widths and mean shifts, whereas the windows centered at $0.465$--$0.470~\mathrm{GeV}$ give a smaller scatter.  This does not mean that the latter windows are new preferred physical regions; it simply shows that the regular part of the numerical zero-contour is cleaner there in the present grid.

%---------------------------------------------------------
\begin{table}[b]\centering
\setlength{\tabcolsep}{9pt}
\begin{tabular}{ccccc}\hline\hline
$M_{+0}^{\rm br}~[\mathrm{GeV}]$ &
$N$ &
$\langle g_{\phi3\pi}^{\rm est}\rangle$ &
median &
$\sigma$\\
\hline
0.450 & 52 & $-22.134$ & $-23.965$ & $8.01$\\
0.455 & 52 & $-22.226$ & $-23.984$ & $6.77$\\
0.460 & 52 & $-23.108$ & $-23.984$ & $5.13$\\
0.465 & 51 & $-23.609$ & $-23.978$ & $1.37$\\
0.470 & 53 & $-23.799$ & $-24.011$ & $0.92$\\
\hline\hline
\end{tabular}
\caption{Sensitivity to the branch-center choice at fixed $\delta M_{+0}=25~\mathrm{MeV}$, $M_{+-}\le0.560~\mathrm{GeV}$, and $\varepsilon_G=100~\mathrm{GeV}^{-7}$. All estimator values are in $\mathrm{GeV}^{-3}$. The median remains stable, while the mean and scatter depend on how many outlier points near unstable quadratic branches enter the selected strip.}
\label{tab:branch_center_sensitivity}
\end{table}
%---------------------------------------------------------

Finally, varying $M_{+-}^{\rm max}$ in the broad range $0.540$--$0.580~\mathrm{GeV}$ leaves the selected sample unchanged.  The reason is kinematic rather than dynamical: after all cuts are imposed, the accepted points in the benchmark branch already lie in the narrower interval
\begin{equation}
0.432\leq M_{+-}\leq0.453~\mathrm{GeV}.
\end{equation}
A meaningful $M_{+-}^{\rm max}$ sensitivity test must therefore be performed near this interval.  The fine scan in Table~\ref{tab:mpmmax_sensitivity} shows that the sample stabilizes once $M_{+-}^{\rm max}\geq0.454~\mathrm{GeV}$.  Below that value, the sample is partially clipped and the mean becomes more sensitive to a small number of accepted points, while the median remains comparatively close to the injected coupling.

%---------------------------------------------------------
\begin{table}[t]\centering
\setlength{\tabcolsep}{9pt}
\begin{tabular}{ccccc}\hline\hline
$M_{+-}^{\rm max}~[\mathrm{GeV}]$ &
$N$ &
$\langle g_{\phi3\pi}^{\rm est}\rangle$ &
median &
$\sigma$\\
\hline
0.438 & 5  & $-18.653$ & $-23.902$ & $15.24$\\
0.442 & 12 & $-20.996$ & $-23.949$ & $10.22$\\
0.446 & 21 & $-21.973$ & $-23.952$ & $7.88$\\
0.450 & 36 & $-22.768$ & $-23.984$ & $6.12$\\
0.454 & 52 & $-23.108$ & $-23.984$ & $5.13$\\
0.458 & 52 & $-23.108$ & $-23.984$ & $5.13$\\
0.462 & 52 & $-23.108$ & $-23.984$ & $5.13$\\
\hline\hline
\end{tabular}
\caption{Sensitivity to the upper cut on $M_{+-}$ at fixed $M_{+0}=0.460\pm0.025~\mathrm{GeV}$ and $\varepsilon_G=100~\mathrm{GeV}^{-7}$. All estimator values are in $\mathrm{GeV}^{-3}$. The accepted branch points occupy $0.432\leq M_{+-}\leq0.453~\mathrm{GeV}$, so the result saturates for $M_{+-}^{\rm max}\geq0.454~\mathrm{GeV}$.}
\label{tab:mpmmax_sensitivity}
\end{table}
%---------------------------------------------------------

Tables~\ref{tab:stability}--\ref{tab:mpmmax_sensitivity}, together with Fig.~\ref{FIG2}, point to the same pattern from three independent angles: the contour-tolerance scan, the branch half-width and center scans, and the $M_{+-}^{\rm max}$ scan. In all of them the median tracks the injected coupling while the mean and width are the first to degrade once the sample drifts toward unstable quadratic roots. We return to what this pattern implies for the method's scope in Sec.~V.

%---------------------------------------------------------
%\section{Data-level extension}
%---------------------------------------------------------
The calculation above is an analytic-density closure test in the FSI-improved amplitude convention.  A direct application to experimental data would require reconstructing $\Delta_G$ from efficiency-corrected Dalitz-bin counts, including the experimental binning, efficiency map, background subtraction, and covariance information.  This is a more delicate operation than a conventional projection comparison, since the zero locus depends on derivatives of the reconstructed density and finite differences amplify bin-by-bin fluctuations.  A data-level implementation should therefore combine efficiency correction with a controlled derivative reconstruction procedure, such as local polynomial fits, spline-based smoothing, or a regularized two-dimensional fit to the Dalitz density prior to differentiation.  The same smoothing prescription must be applied to pseudo-data and real data. Otherwise, the derivative field may acquire a method-dependent bias that is unrelated to the physical direct term.

The stability criteria used above give a practical guide for such an implementation.  The zero-contour should be reconstructed as a geometric object rather than inferred from isolated bins. Resonance-band and $\omega$-region vetoes should be applied before assigning a local estimator value, and the size of the quadratic discriminant, $B_{\rm dir}$, and $C_{\rm dir}$ should be monitored, since these determine whether a local root of Eq.~(\ref{eq:cest}) is numerically meaningful.  In addition, the covariance of neighboring bins should be propagated through the derivative reconstruction, because the zero locus position and the estimator value are correlated observables.  We do not quote pseudo-data closure numbers here. A consistent pseudo-data study should be generated directly from the same FSI-improved amplitude, using the branch $M_{+0}=0.460\pm0.025~\mathrm{GeV}$ and the quadratic estimator of Eq.~(\ref{eq:cest}), and then repeated under realistic binning and efficiency assumptions. Such a study should explicitly vary the expected event sample size to quantify derivative noise, repeat the reconstruction under several Dalitz-bin sizes to test bin-size dependence, and report three closure observables: the reproducibility of the zero-contour position, the bias of the branch-selected median estimator, and the growth of outliers as the tolerance $\varepsilon_G$ is relaxed.

%---------------------------------------------------------
\section{Conclusion}
%---------------------------------------------------------
We have formulated the zero-locus diagnostic for $\phi\to\pi^+\pi^-\pi^0$ using an Omnès-improved reference amplitude. The resonant $\rho\pi$ channels are dressed by the complex Omnès factors, the direct term is dressed by the averaged Omnès factor, and the real constant background is included in the reference amplitude. With this convention, the reduced amplitude can be written as $\mathcal F^{\rm FSI}=R+c e^{i\delta_{\rm dir}}D$, and the zero-locus condition leads to the quadratic local identity in Eq.~(\ref{eq:cest}).

The closure test in Sec.~IV shows that this identity recovers the injected coupling on a low-mass branch around $M_{+0}=0.460\pm0.025~\mathrm{GeV}$, with the median estimator stable under contour-tolerance and branch-window variations while the mean and standard deviation flag the onset of unstable quadratic roots. We do not repeat those numbers here; the point worth stressing is what the scan pattern means rather than what it reproduces. The median's robustness and the mean's fragility are two sides of the same algebraic fact: Eq.~(\ref{eq:cest}) is a ratio of small derivative quantities, so a handful of points near $\mathcal D_c\simeq0$ can dominate a sample average without ever displacing its center. This is precisely the behavior one would want from a diagnostic meant to flag where a derivative-based observable can be trusted, rather than from an estimator meant to stand on its own.

This also clarifies the method's actual scope. The benchmark coupling used throughout is an injected input, and the zero locus is built from a density that already contains it, so the present results demonstrate internal consistency of the FSI-improved convention rather than a measurement of $g_{\phi3\pi}$ in any convention-independent sense. The diagnostic's value lies elsewhere: in pointing to Dalitz-plane regions where a global fit's sensitivity to the direct term, the background, and the FSI treatment can be cross-checked locally, and in giving a concrete, reproducible criterion -- regularity of the zero contour together with the cuts of Eq.~(\ref{eq:eps_dmin_cuts}) -- for choosing such regions in advance of a fit, rather than after one.

Two extensions follow naturally from this scope. First, the present closure test uses an exact analytic density; an experimental application requires propagating efficiency corrections, binning, and bin-to-bin covariance through the derivative reconstruction, since finite differences amplify exactly the kind of local fluctuation that the median/mean contrast above is designed to expose. Second, the normalization mismatch with the KLOE $A_{\rm dir}=a_d e^{i\phi_d}$ parametrization noted in Sec.~I means that any future comparison to data must first fix a common amplitude convention; the zero-locus construction itself does not resolve that mismatch, and should not be read as doing so. A pseudo-data study within the same FSI-improved convention is therefore the natural next step. Such a study should vary the sample size and binning, and should track the reproducibility of the zero-contour position, the bias of the branch-selected median, and the growth of outliers as $\varepsilon_G$ is relaxed.
%---------------------------------------------------------
\acknowledgments
%---------------------------------------------------------
This work was supported by grants from the National Research Foundation of Korea (NRF), funded by the Korean government (MSIT) (RS-2025-16065906, RS-2026-25616630, RS-2026-25611774, and RS-2024-00436392).

%---------------------------------------------------------

%---------------------------------------------------------
\end{document}